\documentstyle[11pt]{article}
\begin{document}
\baselineskip 22pt
\rightline{CU-TP-817}
\vskip -2mm
\rightline{hep-th/9702107}
\vskip 1cm

\centerline{\Large\bf Monopoles and Instantons }
\centerline{\Large\bf on Partially Compactified D-Branes} 
\vskip 1cm
\centerline{\large\it Kimyeong Lee \footnote{e-mail: klee@phys.columbia.edu}
and Piljin Yi \footnote{e-mail: piljin@phys.columbia.edu}}
\vskip 2mm
\centerline{Physics Department, Columbia University, New York, NY 10027}
\vskip 2cm
\centerline{\bf ABSTRACT}

\begin{quote}
Motivated by the recent D-brane constructions of world-volume monopoles and
instantons, we study the supersymmetric $SU(N)$ Yang-Mills theory on 
$S^1\times R^{3+1}$, spontaneously  broken by a Wilson loop.
In addition to the usual $N-1$ fundamental monopoles, 
the $N$-th BPS monopole appears from the Kaluza-Klein sector.
When all $N$ monopoles are present, net magnetic charge vanishes and the
solution can be reinterpreted as a Wilson-loop instanton of unit Pontryagin 
number. The instanton/multi-monopole moduli space is explicitly constructed,
and seen to be identical to a Coulomb phase moduli space of a $U(1)^N$ gauge
theory in $2+1$ dimensions related to Kronheimer's gauge theory of $SU(N)$ 
type. This extends the results by Intriligator and Seiberg to the finite 
couplings that, in the infrared limit of Kronheimer's theory, the Coulomb 
phase parameterizes a centered $SU(N)$ instanton. We also elaborate 
on the case of restored $SU(N)$ symmetry.

\end{quote}

\newpage

\section{Solitons on D-Branes}

Recently there has been a considerable interest in the low energy
D-brane dynamics \cite{D} and its relation to the supersymmetric Yang-Mills
systems. When there are $N$ parallel D$p$-branes, their low energy
dynamics is described by a reduction of the $N=1$
supersymmetric 10-dimensional Yang-Mills system of the gauge group
$U(N)$  to the $p+1$ dimensional superysmmetric Yang-Mills-Higgs
system \cite{Witten}. The physics of the supersymmetric Yang-Mills 
systems can be understood by that of the D-brane dynamics, and vice versa,
enriching our understanding of both subjects.

BPS magnetic monopoles and instanton-like solitons in
supersymmetric Yang-Mills systems have been understood
in the D-brane language recently. The key point is that the Ramond-Ramond (RR) 
charge carried by the D-brane \cite{RR} can be also carried by world-volume
instantons and monopoles. 
The coupling (between the RR gauge field $C_{p+1}$ and the world volume
gauge field) that is responsible for this is succinctly written as
\begin{equation}
\int_{B_p}\sum_{\delta\ge 0} C_{p-2\delta+1}\wedge tr\,e^{{\cal F}/2\pi i}
\end{equation}
where the integral is over the world-volume of D$p$-branes for each $p$
\cite{D}. $\cal F$ is the world-volume Yang-Mills field strength.

For instance, when an open D$(p-2)$-brane ends on a D$p$ brane, the RR charge
carried by the former must be somehow cancelled by a Yang-Mills soliton 
on the D$p$-brane, of co-dimension three \cite{Dbound}. From the above  
coupling, it is clear a soliton of charge
\begin{equation}
\oint_{S^2}{\cal F},
\end{equation}
is a source for $C_{p-1}$ RR gauge field, so a RR charge conservation 
requires that the boundaries of the D$(p-2)$-brane carry the magnetic 
charge with respect to D$p$-brane world-volume gauge field \cite{miao}. 

On the other hand, Yang-Mills instantons appear, say, 
when there are $N$ parallel D4-branes overlapping each other
so that $U(N)$ symmetry is restored.  If a D0-brane approaches the
D4-branes from infinity and touches the D4-branes, it could melt away,
leaving an $SU(N)$ instanton (in real time) on the D4-brane \cite{Douglas}. 
This process conserves the RR charge as both D0 branes and
instantons carry the same charge. The instanton energy is identical to
the D0-brane mass and one can interpret the instanton as the threshold
bound state between a D0-brane and $N$ overlapped D4-branes. This
picture has been studied in many variations connected under the
T-duality. In particular the D-brane configuration has been shown to
be connected to the ADHM construction of the instanton configurations.

In this paper we consider cases where at least one direction of the space-time
is compactified on a circle, so that a D3-brane/D1-brane configuration in 
the type IIB theory is T-dual to a D4-brane/D0-brane configuration in the
type IIA. Both magnetic monopoles and instantons appear in the compactified, 
five-dimensional Yang-Mills system of D4-branes, when the gauge
symmetry is broken by a nontrivial Wilson loop. This will lead to a new 
understanding of magnetic monopoles and instantons in a compactified 
Yang-Mills theory. In the latter half of the paper, we will 
concentrate on the moduli space of such solitons.

\section{The Low-Energy Effective Field Theory} 

We take the spacetime to be $\tilde  S^1\times R^{3+1}\times T^5 $.
Let $N$ parallel D$p$-branes  ($p\ge 3$) overlap with the noncompact part 
$R^{3+1}$. Up to various T-dual transformations, we may as well take $p=3$
so that these D-branes lie entirely along $R^{3+1}$. 
When $N$ D-branes are all separated from each
other, the theory is in the Coulomb phase where it admits  magnetic
monopole solutions. Furthermore, we will assume that all D3-brane positions
are aligned along $\tilde S^1$, so we are effectively working in $\tilde S^1
\times R^{3+1}$. Finally, upon a T-dual transformation, we map $\tilde S^1$
of circumference $\tilde
 L$ to its dual $S^1$ of circumference $L=4\pi^2\alpha'/
\tilde  L$, which is now wrapped around by $N$ D4-branes.

The world-volume theory in question is then five-dimensional ${\cal N}=4$ 
$U(N)=U(1)\times SU(N)$ Yang-Mills compactified on $S^1$ \cite{Witten}. 
The Abelian part $U(1)$ of the gauge group
will be ignored in our discussion here.
The bosonic part of the effective Lagrangian on $S^1\times R^{3+1}$ is
\begin{equation}
{\cal L} =  \mu \int_{S^1\times R^{3+1}}  \, tr \left\{
 - \frac{1}{2}{\cal F}_{MN}{\cal F}^{MN} +  \sum_P
{\cal D}_M\Phi_P {\cal D}^M\Phi_P 
+ \frac{1}{2} \sum_{P,Q}[\Phi_P,\Phi_Q]^2 \right\} 
\end{equation}
The dimensionful coupling $\mu$ is proportional to the D4-brane tension
$\tau_4$, and, for later use, we express it in terms of 
the D-string tension $\tilde\tau_1$,
\begin{equation}
\mu =2\pi^2\alpha'^2\,\tau_4
                          =\frac{\tilde\tau_1 \tilde L}{8\pi^2}.
\end{equation}
The circumference $\tilde L$ of $\tilde S^1$ enters because the
T-duality transformation rescales the string coupling $e^\phi\propto 1/
\tilde\tau_p$. The relevant symmetry breaking is via a Wilson loop along $S^1$,
so we shall subsequently ignore the five scalar fields $\Phi_P$.

The theory admits the BPS bound, as usual, and in the absence of electric
excitation, the energy functional is 
\begin{equation}
{\cal E}=\frac{\mu}{2}\int_{S^1\times R^3}tr {\cal F}_{\mu\nu}{\cal F}^{\mu\nu}
\end{equation}
with the Greek indices ranging from 1 to 4. This is bounded by
\begin{equation}
{\cal E}\ge \frac{\mu}{2}\int_{S^1\times R^3} tr\; \frac{1}{2}
              \epsilon_{\mu\nu\alpha\beta}
              {\cal F}_{\mu\nu} {\cal F}^{\alpha\beta}
         = 8\pi^2 \mu\, p_1({\cal F})
\end{equation}
where $p_1({\cal F})$ is the Pontryagin number of the Yang-Mills field,
\begin{equation}
p_1({\cal F})\equiv\frac{1}{8\pi^2}\int_{S^1\times R^3} tr {\cal F},
\wedge{\cal F}.
\end{equation}
and the bound is saturated when the Yang-Mills field solves the BPS equation
\begin{equation}
{\cal F}_{\mu\nu}=\frac{1}{2}\epsilon_{\mu\nu\rho\sigma}{\cal F}^{\rho\sigma},
\end{equation}
on $S^1\times R^3$.

\section{Fundamental Monopoles and Wilson-Loop Instantons}

Splitting the gauge field
${\cal A}_\mu$ into the noncompact part $A_i, i=1,2,3$ and the compact 
part ${\cal A}_4$, the BPS equation is
\begin{equation}
B_i=D_i{\cal A}_4-\partial_4 A_i.
\end{equation}
With this notation the BPS bound can be written as,
\begin{equation}
{\cal E}\ge 8\pi^2\mu\,
p_1({\cal F})= 2\mu\int_{S^1}\int_{R^3}  tr \left[B_i D_i{\cal A}_4
-\frac{1}{2}\epsilon_{ijk}A_j\partial_4 A_k\right] .
\end{equation}
We first  observe that all three-dimensional BPS monopole solutions \cite{bps}
are also solutions of this equation just by setting $A_i$ independent of the 
periodic coordinates $x^4$ and regarding ${\cal A}_4$ as an adjoint
scalar field. Thus, this theory admits the $N-1$ spherically symmetric, 
fundamental monopole solutions \cite{erick}, each carrying a 
distinct topological winding number in $\pi_2(SU(N)/U(1)^{N-1})=Z^{N-1}$.
Each of these fundamental solutions is characterized for having only 4
zero modes.

However, there are other  solitonic solutions of the
same monopole charges, an infinite number of them as a matter of fact.
To see this, it suffices to recall that the BPS solution exists for generic
asymptotic value ${\cal A}_4(\infty)$. Because ${\cal A}_4$ is in 
fact a component of the gauge connection along the compact direction, it
can be shifted under a large gauge transformation as
\begin{equation}
{\cal A}_4 \rightarrow {\cal A}_4 +\Delta
\end{equation}
for some constant hermitian matrix $\Delta$. Monopole solutions of larger 
masses are obtained
by first solving the three-dimensional BPS equation with the boundary condition
${\cal A}_4(\infty)=\langle{\cal A}_4 \rangle+n\Delta$   and then
performing a large gauge transformation back to ${\cal A}_4(\infty)=
\langle{\cal A}_4\rangle$. 
This works for $n\ge 0$ whenever $tr \,(\langle{\cal A}_4 \rangle\,\Delta)>0$
and $e^{iL\Delta}$ belongs to the center of the gauge.

Consider the simplest case of $SU(2)$ for $N=2$. We write the generic Wilson 
loop as
\begin{equation}
\langle{\cal A}_4\rangle =\frac{2\pi\eta}{L} \hat Q
\end{equation}
for $0\le\eta<1$ and where $\hat Q$ is the unbroken $U(1)$ charge operator
normalized to be unit for the massive vector meson. In the unitary gauge,
$\hat Q$ would be a diagonal matrix $diag(1/2,-1/2)$. 
The usual BPS monopole is then
a solitonic solution where ${\cal A}_4$ interpolates between 0 at origin
and $2\pi\eta\,\hat Q/L$ at asymptotic infinity. The relevant $\Delta$ can
be chosen as
\begin{equation}
\Delta=\frac{2\pi}{L}\hat Q ,
\end{equation}
so the infinite tower of monopole solutions simply correspond to rescaled
spherically symmetric BPS solutions that allow ${\cal A}_4$ to interpolate 
between 0 and $2\pi(\eta+k)\,\hat Q/L $ for any nonnegative integer $k\ge 0$.
After a large gauge transformation, we may keep ${\cal A}_4(\infty)$ at
$2\pi\eta\,\hat Q/L$ but let ${\cal A}_4(0)= -2\pi k\,\hat Q/L$. (As long
as the Wilson loop at origin is invariant under the global gauge rotations,
such a solution is perfectly regular and acceptable.)

The masses of these monopoles are easy to evaluate. The monopole solution 
is independent of $x^4$ when we choose ${\cal A}_4(0)=0$, and then its BPS 
mass formula is similar to that of 3-dimensional Yang-Mills theory \cite{bps},
\begin{equation}
{\cal E}\ge 8\pi^2\mu\,p_1({\cal F})= 2\mu\int_{S^1}\int_{R^3}  tr 
\left[B_i D_i{\cal A}_4\right]=8\pi^2\mu\,(k+\eta) .
\end{equation}
Note that the monopoles also carry a Pontryagin number $k+\eta$.

This is a perfectly sensible result from the D-brane perspective. There,
a fundamental monopole is simply a D-string segment that stretches between an
adjacent pair of D3-branes along $\tilde S^1$ \cite{miao}. The D3-brane 
position on $\tilde S^1$ is dictated by the Wilson loop. For $SU(2)$, their 
coordinate $\tilde x^4 \:(\in [-\tilde L/2,\tilde L/2])$ is given by 
$\pm \eta \tilde L/2$, as can be deduced from the fact that both $\eta=0$
and $\eta=1$ correspond to a symmetric phase where the two D3-branes must
coincide. A D-string of length $\eta\tilde L$ stretching between the two 
D3-branes carries a fundamental magnetic charge, and is 
naturally identified with the fundamental monopole above ($k=0$). Note that
the mass formula obtained from the field theory does agree with the
D-brane picture. The mass is 
\begin{equation}
{\cal E}\ge 8\pi^2\mu\eta=8\pi^2\,\frac{\tilde\tau_1\tilde L}{8\pi^2}\,
\eta=\tilde\tau_1
\,(\eta\tilde L) .
\end{equation}
which is simply the D-string tension multiplied by the length. 

In addition, we may have a closed D-string wrapping around $\tilde S^1$. Such 
a loop carries no magnetic charge and is of mass $\tilde\tau_1\tilde L=
8\pi^2\mu$. The configuratons of a single segment of length $\eta\tilde L$ 
combined with 
$k$ loops of such closed D-strings has exactly the right mass and charge to 
form the infinite tower of monopoles obtained above. What is the field theory
soliton that corresponds to the closed D-string loop? The answer is obvious
once we made the above identification. Removing the fundamental monopole to 
asymptotic infinity, one obtains a $k$ closed loops of D-strings of 
zero magnetic charge. Its Pontryagin number $p_1({\cal F})$ is $k$,
so each loops of D-string must be realized as an
$SU(2)$ instanton on $S^1\times R^3$. 

The same reasoning goes through for $SU(N)$. Writing the expectation value
$\langle {\cal A}_4 \rangle$ in a unitary gauge,
\begin{equation}
\langle {\cal A}_4 \rangle=diag(\eta_1,\eta_2,\eta_3,\dots,\eta_N)\,
\frac{2\pi}{L}, \qquad \frac{1}{2}\ge \eta_1 >\eta_2>\cdots>\eta_N\ge 
-\frac{1}{2}
\end{equation}
the $N-1$ fundamental monopoles are of masses $8\pi^2\mu\,(\eta_a-\eta_{a+1})$,
as can be seen from an $SU(2)$ embedding. The infinite tower of monopoles 
for each fundamental charge can be treated by a simple $SU(2)$ embedding, 
and this again results in the mass formula,
\begin{equation}
{\cal E}\ge 8\pi^2\mu p_1({\cal F})=8\pi^2\mu\, (k+\eta_{a}-\eta_{a+1})
\label{mass}
\end{equation}
Thus the higher mass monopole is again interpretable as a combination of
a fundamental
monopole and $k$ $SU(N)$ instantons. On the D-brane side, the $N$ D3-branes 
are at $\tilde x^4=\eta_i\tilde L$, so the string segments bewteen adjacent 
pairs are of masses $\tilde\tau_1\,(\eta_{a}-\eta_{a+1})\tilde L =8\pi^2\mu\,
(\eta_{a}-\eta_{a+1})$ with the fundamental charge, while the chargeless, 
closed loop of D-brane has the mass $\tilde\tau_1\tilde L=8\pi^2\mu$. 
Comparing this to Eq.~(\ref{mass}), it is pretty clear that the closed loop of
D-string corresponds to an instanton of $p_1({\cal F})=1$.

This is in accordance with the fact that,
upon the T-duality transformation from $\tilde S^1$ to $S^1$, a single 
D-string loop crossing $N$ D3-branes turn into a D0-brane on 
$N$ D4-brane, a natural candidate for an $SU(N)$ instanton. 
Because the instantons exist even when the gauge symmetry is broken by a
Wilson loop, we shall call them the Wilson-loop instantons.

\section{The Wilson-loop Instanton as $N$ Fundamental Monopoles}

Such a closed loop of D-string can break up into segments between adjacent
pairs of D3-branes. With D3-branes separated along $\tilde S^1$, there are
exactly $N$ such monopole-like segments. However, so far we have isolated
only $N-1$ species of fundamental monopoles, each represented by a
three-dimensional spherically symmetric solution. We need one
more monopole solution in order to match the D-brane picture.
In the D-brane picture, this corresponds to a segment stretching between
the first and the $N$-th D3-branes directly, and clearly is possible thanks
to the compact nature of $\tilde S^1$. On the other hand, 
since a closed loop of D-string carries no magnetic charge 
with respect to the unbroken $U(1)^{N-1}$, the $N$-th monopole must have a 
charge opposite to the sum of the other $N-1$ fundamental monopoles.

To see how a BPS solution of a ``wrong'' magnetic charge arises in the 
Yang-Mills theory, let us again consider the simplest case of $SU(2)$.
Here, it is also useful to recall why a pair of BPS monopoles
has no static force between them: it is because the vector force is 
canceled by that of the scalar. In order to flip the magnetic charge
and still solve the BPS equation, we need to make the absolute 
value of ${\cal A}_4$ to increase (rather than decrease)
toward origin. With ${\cal A}_4(\infty)= 2\pi\eta\hat Q/L$ as above,
possible choices for ${\cal A}_4(0)$ are $(k+1)2\pi \hat Q/L$ for any 
nonnegative integer $k$.

How do we know a BPS solution wth such boundary conditions exists? Because
such a configuration can be gauge-transformed to the usual 
BPS solution through a large gauge transformation that shifts
${\cal A}_4\rightarrow {\cal A}_4-(k+1)2\pi \hat Q/L$ everywhere, followed
by a global gauge rotation $\hat Q\rightarrow -\hat Q$. The resulting
configuration is that of the ordinary BPS solution whose ${\cal A}_4$
interpolate between $0$ at origin to $(k+1-\eta)2\pi\hat Q/L$ at
asymptotic infinity. The mass is again easily estimated in the latter gauge, 
and 
\begin{equation}
{\cal E}\ge 8\pi^2\mu^2\,(k+1-\eta).
\end{equation}
The infinite tower of monopoles given by this is a precise analog of 
the infinite towers we encountered above. The minimal case of $k=0$ is again
a fundamental monopole in that it carries precisely 4 zero modes.
Generalization to $SU(N)$ proceeds similarly.

This second (or  the $N$-th) fundamental monopole is possible because the 
$x^4$ direction is compact. In the gauge where the ordinary fundamental 
monopoles are $x^4$-invariant,
this solution must have a Kaluza-Klein momentum along $x^4$.
In reality, however, both (or all $N$ for $SU(N)$) monopoles are on equal 
footing, since one can always perform a large 
gauge transformation to get rid of the $x^4$-dependence of the second 
(or the $N$-th) fundamental monopole, as we saw above.

We need one more field theory computation to complete the picture. If the
higher mass monopole is indeed a mere sum of a fundamental monopole and 
$k$ instantons, and if the instanton itself is a sum of $N$ monopoles, the
zero-mode counting must reflect this. In particular, the higher mass monopoles
of fundamental charge must carry a large number of zero modes which should
be also consistent with that of an instanton.

Following Brown et. al. \cite{Brown}, we  write the zero mode equation as,
\begin{equation}
\sigma_\mu {\cal  D}_\mu \Psi=0,
\end{equation}
where ${\cal D}_\mu$ is the background covariant derivative and $\Psi\equiv
\bar\sigma_\mu \delta{\cal A}_\mu$ satisfies a reality constraint coming
from the fact that the  zero modes $\delta A_\mu$ are hermitian matrices. 
The $2\times 2$ matrices $\sigma_\mu$ are given by  $(\sigma_j, i)$ while 
$\bar\sigma_\mu=(\sigma_j,- i)$. 

Now consider the infinite tower of monopoles of the fundamental charge in
the $SU(2)$ theory. Performing a Fourier expansion of $\Psi$ with respect 
to the internal periodic coordinate $x^4$,
\begin{equation}
\Psi=\sum_{m\in Z} e^{i 2\pi m x^4/L}\Psi_{(m)},
\end{equation}
the zero-mode equation reduces to a three-dimensional one due to 
E. Weinberg \cite{erick}, now with a bare mass,
\begin{equation}
\sigma_\mu  D_\mu \Psi_{(m)}=\frac{2\pi m}{L}\Psi_{(m)},
\end{equation}
where $D_j\equiv{\cal D}_j$ and $D_4\equiv{\cal D}_4-\partial_4$. From 
Ref.~\cite{erick},
one can easily see that this admits 4 normalizable solutions whenever the 
bare mass term is smaller than the scale set by the monopole mass, i.e.,
$|m|<\eta+k$. Since $\eta >0$ in the broken phase, 
$2k+1$ Fourier modes contribute 4 each, and
\begin{equation}
8k+4 \qquad 
\end{equation}
is the total number of zero modes.
When we consider the $SU(N)$ monopoles, the same reasoning goes through since
we may obtain the necessary BPS solutions by embedding the $SU(2)$ solutions.
One difference is that there are also some ($SU(N)$) zero modes that transform 
as doublets with respect to this embedded $SU(2)$. After this is properly 
taken into account, we find
\begin{equation}
4Nk+4
\end{equation}
as the total number of zero modes.

This zero-mode counting is clearly consistent with the interpretation that 
the higher mass monopole is a fundamental monopole of the same charge 
combined with a chargeless collection of $Nk$ fundamental monopoles 
of 4 zero modes each. In turn, the index theorem applied to the Wilson-loop 
instantons of $p_1({\cal F})=k$ gives the bulk contribution,
\begin{equation}
2\times \frac{1}{8\pi^2}\int_{S^1\times R^3} Tr{\cal F}\wedge{\cal F}
=4N \times \frac{1}{8\pi^2}\int_{S^1\times R^3} tr{\cal F}\wedge{\cal F}=4Nk
\end{equation}
while the boundary contribution is expected to be null for integral 
$p_1({\cal F})$
as in $R^4$ case. The first trace $Tr$ is over the $SU(N)$ adjoint 
representation, and we used the identity $Tr(\cdots)=2N\times tr(\cdots)$ 
for $SU(N)$. Again the zero-mode counting is consistent with the above
picture that the $Nk$ monopoles are in fact $k$ $SU(N)$ instantons.  
The interpretation of a Wilson-loop instanton as $N$ fundamental monopoles 
is thus complete in the purely Yang-Mills theory context.

\section{The Exact Moduli Space and a 3-D Gauge Theory}

A simple consequence is that the one-instanton moduli space is identical
to that of $N$ fundamental  monopoles. When the $N$ monopoles are   
well-separated, we can infer the approximate form of the metric from their
long-range interaction. Following Gibbons and Manton \cite{Manton},  
Lee, Weinberg and Yi constructed the general form of such an approximate 
metric \cite{klee1}. Applied to the present case, it  gives
\begin{equation}
{\cal G}=M_{ab}d{\bf x}_a\cdot d{\bf x}_b+
(M^{-1})_{ab}(d\xi_a+{\bf W}_{ac}\cdot d{\bf x}_c)(d\xi_b+{\bf W}_{bd}
\cdot d{\bf x}_d) ,\label{metric}
\end{equation}
where the diagonal components of $N\times N$ matrix $M$ are
\begin{eqnarray}
M_{aa}&=& m_1+\frac{1}{|{\bf x}_1-{\bf x}_{N}|}+
        \frac{1}{|{\bf x}_1-{\bf x}_{2}|}  \qquad a=1, \nonumber\\
M_{aa}&=& m_a+\frac{1}{|{\bf x}_a-{\bf x}_{a-1}|}+
        \frac{1}{|{\bf x}_a-{\bf x}_{a+1}|}  \qquad a=2,\dots,N-1, \nonumber\\
M_{aa}&=& m_N+\frac{1}{|{\bf x}_N-{\bf x}_{N-1}|}+
         \frac{1}{|{\bf x}_N-{\bf x}_{1}|}  \qquad a=N,
\end{eqnarray}
with  $m_a$ being the (rescaled) mass of the $a$-th monopole, which is
located at ${\bf x}_a$ in $R^3$, and the only nonvanishing off-diagonal 
components are
\begin{equation}
M_{1N}=M_{N1}=-\frac{1}{|{\bf x}_1-{\bf x}_{N}|},\qquad
M_{ab}=-\frac{1}{|{\bf x}_a-{\bf x}_{b}|}\quad (|a-b|=1).
\end{equation}
The vector  potential $\bf W$ is related to the the scalar potential $M$
by
\begin{equation}
\nabla_c M_{ab}=\nabla_c \times {\bf W}_{ab},
\end{equation}
which ensures that the metric is hyper-K\"ahler. The $U(1)$ 
coordinate $\xi_a/2$ for each $a$ is periodic in $2\pi$ and gives rise to 
integer-quantized dyonic excitation of the $i$-th monopole.

If the $N$-th monopole is absent (or infinitely far away from the other
$N-1$ monopoles), such an approximate metric is known to be exact 
\cite{klee1,Chalmers}. One compelling physical reason behind this is the 
fact that the $N-1$ unbroken $U(1)$ gauge symmetry 
prohibits certain short-distance corrections that allow an electric charge
transfer among the $N-1$ fundamental monopoles. With the addition of the 
$N$-th monopole, there are still only $N-1$ $U(1)$ symmetries from the 
original gauge group, 
so it may appear that a short distance correction is inevitable. 

However, there is another $U(1)$ symmetry, namely the translation along
$S^1$, which acts to preserve an electric charge. One easy way of seeing this 
is again from the D-brane picture. An electric charge is carried by open 
fundamental string segments stretching between D3-branes. When all $N$
monopoles carry the same (absolute) quantized amount of electric charges,
the situation is that of a closed string winding around $\tilde S^1$.
Upon a T-dual to $S^1$, this winding number is translated to the 
conserved momentum along $S^1$. From  the low energy perspective, we can 
also compute the momentum $P_4$
\begin{equation}
P_4\propto \int_{S^1\times R^3}tr\,{\cal  F}_{0\mu}{\cal  F}_{\mu 4},
\end{equation}
which, for $N=2$ and $k=1$, evaluates to
\begin{equation}
P_4\propto q_1\eta+q_2(1-\eta) =q_2+\eta (q_1-q_2),
\end{equation}
when $q_1$ and $q_2$ are the electric charges (our convention 
is such that the total electric gauge charges are zero when $q_1=q_2$) 
on the first and the second fundamental $SU(2)$ monopoles
respectively. The translation invariance along $x^4$ thus
preserves a linear combination of the two electric charges.\footnote{$P_4$
is not properly quantized because it is not the Noether momentum. The latter
is found by dropping a $\eta$-dependent surface term that is independently
conserved.}

The $N$ independent conserved electric charges along with the hyper-K\"ahler
property of the moduli space implies that the asymptotic form above must
be in fact the exact expression. One should be able to set up an argument 
similar to those in Ref.~\cite{Chalmers} and show this explicitly. 
We have obtained the instanton moduli space of an $SU(N)$ instanton on 
$S^1\times R^3$ for an arbitrary Wilson loop from the equivalent 
multi-monopole configuration.

The same form of metric has recently appeared in a work by Intriligator and 
Seiberg \cite{2+1}, 
as that of the Coulomb branch moduli space of a $U(1)^N$ gauge
theory with four extended supersymmetry in 
$2+1$-dimension.\footnote{Intriligator and Seiberg actually 
considered a $U(1)^N/U(1)$ gauge theory due to Kronheimer. But the difference 
is simply that of whether or not one factors out a trivial part $S^1\times 
R^3$ of the moduli space.} The theory has $N$ species of electrons of charge 
$(1,-1)$ with respect to each adjacent pairs of $U(1)$ gauge groups, and
the two moduli space coincides if the bare masses of the electrons vanish.

This identity can be understood by adapting the method of Hanany and Witten
\cite{Hanany}.
We consider $N$ parallel NS 5-branes separated along a circle $\tilde S^1$.
We will take the rest of the space-time to be noncompact $R^{8+1}$.
Also put one 3-brane segment between each adjacent pairs of the 5-branes.
The corresponding solitons of co-dimension three in the 5-brane world-volume 
theory are precisely the monopoles we discussed above, or collectively an
$SU(N)$ instanton, through a series of S- and T- dualities as well as some 
decompactifications. 

On the other
hand, Hanany and Witten also identified the effective $2+1$-dimensional  
theory on such 3-brane segments, and the rule is that each segment produces 
a $U(1)$ vector multiplet and each adjacent pair of segments gives a 
hypermultiplet of charge $(1,-1)$ with respects to the two $U(1)$'s. With 
the $N$ 3-brane segments parallel to $\tilde S^1$, then, the gauge theory is 
$U(1)^N$ with $N$ species of electrons linking pairs of $U(1)$'s successively.
The Coulomb phase of this theory is parameterized by the 3-brane 
configurations, which are nothing but the instanton/multi-monopole 
configurations from the 5-brane perspective. The two moduli spaces are
identical, and the three-dimensional $U(1)^N$ couplings $g_a$ are 
determined by the monopole masses,
\begin{equation}
\frac{1}{g_a^2}\sim m_a .
\end{equation}
Furthermore, our assertion that the metric written above is exact, is
reflected in the fact that the Coulomb phase moduli 
space metric of the Abelian $U(1)^N$ theory receives no nonperturbative
correction. 

One particular linear combination of the $U(1)$ gauge fields
is free, and if it is removed, we recover Kronheimer's theory of $SU(N)$ 
type \cite{2+1,Kron}. On the other side of the correspondence, this has the 
effect of 
factoring out the center-of-mass motion on $S^1\times R^3$, so the 
Coulomb phase of the Kronheimer theory coincides with the relative part
of the instanton/multi-monopole moduli space.

A special case of this result was anticipated by Intriligator and Seiberg 
\cite{2+1}. They noted that the infrared limit of the Kronheimer theory of 
type $SU(N)$ is ``mirror'' to a $U(1)$ theory with $N$ electrons. The Higgs 
phase of the latter had been interpreted as the moduli space of an instanton 
located at a fixed point in $R^4$, and, under the proposed mirror symmetry, 
should be mapped to the Coulomb phase of the infinite coupling Kronheimer 
theory. Such a mirror mapping was subsequently justified by various
authors \cite{Hanany,others}.
In view of the relationship between the monopole masses $m_a$ and the $U(1)^N$
couplings $g_a$, this can be seen to be a special case of our result
in the limit of decompactified $S^1$ (dual to vanishing $\tilde S^1$).

\section{Symmetric Phase and the Calabi Metric}

When the Wilson loop becomes trivial, the $SU(N)$ gauge symmetry is restored
and the moduli space must reduce to that of a symmetric-phase instanton 
on $S^1\times R^3$, also known as the periodic instanton.
In this limit, the first $N-1$ monopoles are massless ($m_1=\cdots
=m_{N-1}=0$). Redefining the position coordinates by ${\bf r}_A={\bf x}_A
-{\bf x}_{A+1}$ for $A=1,\dots,N-1$ and ${\bf R}={\bf x}_N$, and the $U(1)$ 
phases by $\psi_A=\sum_{a=1}^A\xi_a$ and $\chi=\sum_{a=1}^N\xi_a$, the 
moduli space metric can be rewritten as
\begin{eqnarray}
{\cal G}&=& m_N \,d{\bf R}^2+\frac{1}{m_N}\,d{\chi}^2 +{\cal G}_{\rm rel},
\nonumber \\
{\cal G}_{\rm rel}&=& C_{AB}d{\bf r}_A \cdot d{\bf r}_B +(C^{-1})_{AB}
            (d\psi_A+{\bf w}_{AC}\cdot d{\bf r}_C)
            (d\psi_B+{\bf w}_{BD}\cdot d{\bf r}_D),
\end{eqnarray}
where the $(N-1)\times (N-1)$ scalar potentials $C_{AB}$ are
\begin{equation}
C_{AA}=\frac{1}{|{\bf r}_A|}+\frac{1}{|\sum_A {\bf r}_A|},\qquad
C_{AB}=\frac{1}{|\sum_A {\bf r}_A|} \quad (A\neq B).
\end{equation}
The  vector potentials satisfy $\nabla_C\times {\bf w}_{AB}=\nabla_C C_{AB}$,
and the $U(1)$ coordinates $\psi_A$ are all of period $4\pi$. 
Note that the metric ${\cal G}_{\rm  rel}$ is devoid of any mass scale.
The relative moduli space described by ${\cal G}_{\rm  rel}$ is thus
valid for any size of $S^1$, and can be considered the moduli space of a 
symmetric-phase $SU(N)$ instanton located at a point in either 
$S^1\times R^3$ or $R^4$. It is precisely the Coulomb phase moduli space
of the infinite coupling Kronheimer theory of $SU(N)$ 
type.\footnote{The coordinates Intriligator and Seiberg 
used in Ref.~\cite{2+1} are 
more like the ${\bf x}_a$'s and the $\xi_a$'s above than the proper relative 
coordinate ${\bf r}_A$'s and $\psi_A$'s, so one must be 
careful to express one of them, say for $a=N$, as functions of those for 
$a=1,\dots,N-1$.} 

The metric ${\cal G}_{\rm  rel}$ itself is a degenerate limit of the 
so-called Calabi metric \cite{Calabi}
which is an $SU(N)$-invariant hyper-K\"ahler metric. 
For ${\cal G}_{\rm rel}$, the $SU(N)$ isometry is clearly related to 
the restored $SU(N)$ gauge symmetry. The relative moduli space in the 
symmetric phase parameterize the gauge orientation of the instanton
beside its size: the principal $SU(N)$ orbit of the Calabi manifold
is $SU(N)/U(N-2)$, and the remaining single coordinate must correspond 
to the instanton size.

The metric ${\cal G}_{\rm  rel}$ possesses an isolated singularity at origin.
For $N=2$, this is particularly easy to see because the Calabi metric is simply
that of the Eguchi-Hanson gravitational instanton whose degenerate limit
is $R^4/Z_2=R^+\times SU(2)/Z_2$. The isolated singularity at origin persists
as we break the $SU(N)$ gauge symmetry by a Wilson loop, because the 
singularity occurs at vanishing instanton size where the scale of the Wilson
loop is negligible. Again, this can be seen explicitly for $N=2$:
the relative moduli space of a pair of  distinct monopoles is always
given by Taub-NUT space locally \cite{Taub}, and thus  by continuity
it has to be a $Z_2$ orbifold of the Taub-NUT space. The massless limit
of the Taub-NUT is $R^4$.

This massless limit of the instanton/multi-monopole moduli space provide 
us with an interesting explanation of a phenomenon found by Rossi in the late
70's \cite{Rossi}. Start with 't Hooft's $SU(2)$ multi-instanton solution 
and line them up along a fixed axis, say $x^4$, at equal distances. The 
resulting configuration is periodic along $x^4$, and effectively a single 
symmetric-phase instanton on $S^1\times R^3$. As usual, there is a single 
moduli that parameterize the instanton size, say $\rho$, in addition to the 
moduli that arise from broken global symmetries. Then,  it was
observed that, as $\rho$ is sent to $\infty$, the periodic instanton 
solution of ever-increasing size approached the usual BPS monopole 
solution in $R^3$ up to a large gauge transformation.

In our picture, the periodic  $SU(2)$ instanton is composed of a pair
of distinct fundamental monopoles. In the limit of restored
$SU(2)$ gauge symmetry, the  Wilson loop is trivial ($\eta=0$) so that the
first fundamental monopole is massless. But the second is still
massive. The situation is reminiscent of those in Ref.~\cite{klee2}; 
as the non-Abelian gauge symmetry is restored, some monopoles become massless
and dissolve 
into a charge cloud that shields the (non-Abelian) magnetic charge of the 
remaining massive monopole. At such a massless limit, some of the
collective coordinates acquire new physical significance, and in particular,
what used to be the inter-monopole distance translates into the size of the 
cloud.

More generally, when we have the $SU(N)$ gauge group restored, $N-1$ of the
monopoles are now massless, and only one, say the $N$-th, remains massive. 
There is again a single collective coordinate $\rho$ that parameterizes the 
cloud size or equivalently the instanton size. In terms of the 
three-dimensional coordinates above, $\rho$ can be redefined to 
satisfy a simple relation
\begin{equation}
\rho^2/L=
|{\bf x}_1-{\bf x}_{2}|+|{\bf x}_2-{\bf x}_{3}|+\cdots+
|{\bf x}_{N-1}-{\bf x}_{N}|+|{\bf x}_N-{\bf x}_1|,
\end{equation}
which can be deduced from the study of the moduli space of two distinct 
massive monopoles in Ref.~\cite{klee2}
that arise upon $SU(N+2)\rightarrow SU(N)\times U(1)^2$. The 
present moduli space results from the latter by putting the two 
massive monopoles at the same point and identifying their electric charges.

Thus the large instanton limit ($\rho\rightarrow \infty$) is realized if at 
least one of the to-be-massless monopoles is removed to the asymptotic
infinity. In fact, by an $SU(N)$ gauge rotation, this is equivalent to
letting ${\bf x}_a=\infty$ for $a=1,\cdots,N-1$ simultaneously. 
Left behind is a single massive monopole at ${\bf x}_N$, which 
certainly can be gauge-transformed to a canonical BPS monopole solution.

The smooth interpolation between the monopole picture and the instanton
picture also tells us something about the multi-monopole configurations 
in the broken phase. Far away from each other, individual
monopole has a clear identity as magnetic solitons on
$R^3$. As their separations grow smaller, however, the size of the 
internal $S^1$ becomes appreciable, and they cannot retain the character of
solitons on $R^3$. Rather, as the length scale progressively 
decreases, the compact nature of $S^1$ will no longer be important and 
they must clump together at a  point in $S^1\times R^3$, 
and look a lot like very small $R^4$ instanton. This is a radical 
departure from what one would expect from ordinary three-dimensional
magnetic solitons, and must be responsible for the isolated
singularity of the relative moduli space at origin.

\section{Conclusion}

We have studied field theory aspects of monopoles and instantons on D-branes
in a compactified spacetime, and found a consistent picture emerging 
from a purely field theoretical perspective. For the gauge group
$SU(N)$, we also found the exact moduli space of a single Wilson-loop
instanton by 
interpreting it as a collection of $N$ distinct fundamental monopoles. 
The relative part of this moduli space is subsequently identified with the 
Coulomb phase moduli space of the three-dimensional Kronheimner theory 
of type $SU(N)$. In the limit where $SU(N)$ is restored, or in 
the infrared limit of the Kronheimer theory, the relative moduli space 
turns out to be the degenerate limit of the Calabi manifold.

There are many directions to explore further. First of all, one may
consider the $S$-duality of the type IIB theory, and look for
threshold bound states. Since our instanton moduli space has the
maximal triholomorphic Abelian symmetry, we suspect that the generalization 
of Gibbons construction \cite{Gibbons} generates threshold 
bound states.

Second, there is a generalized Nahm formalism \cite{Nahm}
in constructing self-dual
solutions on $S^1\times R^3$ such as our monopoles and instantons
\cite{Nahm5}. 
The basic aspect has been explored before and it should be interesting 
to understand it further in the context of D-brane physics.

Third, we may consider the Yang-Mills theory on compact D4-branes, such as
on a four-torus. Instanton should persist but the concept of magnetic monopole
is no longer available. One outstanding question is how to construct the 
moduli space in such cases. It would be most interesting to see if a simple
derivation as ours is also possible. 

\vskip 5mm

P.Y. thanks J. Maldacena for a stimulating  conversation and  also  for
drawing his attention to compactified Yang-Mills
systems. The authors are also grateful to A. Dancer and A. Swann for making
their manuscript available prior to its publication. K.L. supported
by the Presidential Young Investigator Fellowship. This work is
supported in part by U.S.  Department of Energy.


\begin{thebibliography}{99}

\bibitem{D}
J. Polchinski, {\it TASI lectures on D-branes}, hep-th/9611050;
references therein.

\bibitem{Witten}
E. Witten, {\it Bound states of strings and p-branes},
Nucl. Phys. B460, 335 (1996), hep-th/9510135.


\bibitem{RR}
J. Polchinski, {\it Dirichlet
branes and Ramond-Ramond charges}, Phys. Rev. Lett. 75, 4724 (1995),
hep-th/9510017.



\bibitem{Dbound} 
A. Strominger, {\it Open p-branes}, Phys. Lett. B383,
44 (1996), hep-th/9512059; M.B. Green and M. Gutperle, {\it Comments
on three-branes}, Phys. Lett. B377, 28 (1996), hep-th/9604091;
A.A. Tseytlin, {\it Self-duality of the Born-Infeld action and
Dirichlet 3-branes of type IIB superstring theory},
Nucl. Phys. B469, 	51 (1996), hep-th/9602064.

\bibitem{miao} 
D.-E. Diaconescu, {\it D-branes, monopoles and Nahm
equations}, hep-th/9608163;
M.R. Douglas and M. Li, {\it D-brane realization of N=2 
super Yang-Mills theory in four dimensions}, hep-th/9604041.



% Dbrane and  instanton
\bibitem{Douglas} 
M.R. Douglas, {\it Gauge fields and D-branes},
hep-th/9604198; E.Witten, {\it Small instantons in
string theory}, Nucl. Phys. B460, 541 (1996), hep-th/9511030; 
M.R. Douglas and G. Moore, {\it D-branes, quivers, and
ALE instantons}, hep-th/9603167.







%the BPS bound
\bibitem{bps}{E.B. Bogomol'nyi, Sov. J. Nucl. Phys. 
24, 449 (1976); M.K. Prasad and C.M. Sommerfield, Phys. Rev. Lett. 
35, 760 (1975); S. Coleman, S. Parke, A. Neveu and C.M. Sommerfield,
Phys. Rev. D15, 544 (1977).}

\bibitem{erick}
E.J. Weinberg, {\it Fundamental monopoles in theories with arbitrary symmetry
breaking}, Nucl. Phys. B203, 445 (1982).

\bibitem{Brown}{L.S. Brown, R.D. Carlitz, and C. Lee,
{\it Massless excitations in pseudoparticle fields,}
Phys. Rev. D{\bf 16}, 417 (1977).} 

\bibitem{Manton}
G.W. Gibbons and N.S. Manton, {\it The moduli space metric for
well-separated BPS monopoles}, Phys. Lett. B356, 32 (1995),
hep-th/9506052. 



\bibitem{klee1}
K. Lee, E.J. Weinberg and P. Yi, {\it The moduli space of many BPS
monopoles for arbitary gauge groups}, Phys. Rev. D 54, 1633 (1996),
hep-th/9602167.

%murray and chalmers
\bibitem{Chalmers}
M.K. Murray, {\it A note on the (1,1,...,1) monopole metric},
hep-th/9605054;
G. Chalmers, {\it Multi-monopole moduli spaces for SU(N) gauge group}, 
hep-th/9605182.



%3-dim gauge theory 
\bibitem{2+1} K. Intriligator and N. Seiberg, {\it Mirror symmetry in three
dimensional gauge theories}, Phys. lett. B387, 513 (1996),
hep-th/99607207.

\bibitem{Hanany}
A. Hanany and E. Witten, {\it Type IIB superstrings, BPS monopoles, 
and three-dimensional gauge dynamics}, hep-th/9611230.

\bibitem{Kron}
P.B. Kronheimer, {\it The  contruction of A-L-E spaces as hyper-K\"ahler 
quotient}, J. Diff. Geom. 29, 665-683 (1989). 

\bibitem{others}
J. de Boer, H. Kori, H. Ooguri, and Y. Oz. {\it Mirror symmetry 
in three-dimensional theories, quivers and D-branes}, hep-th/9611063;
M. Porrati and A. Zaffaroni, {\it M-theory origin of mirror symmetry in 
three-dimensional gauge theories}, hep-th/9611201.




%Calabi
\bibitem{Calabi} G.W. Gibbons and P. Rychenkova, {\it Hyper-K\"ahler
quotient construction of BPS monopole moduli spaces}, hep-th/9608085;
A. Dancer and A. Swann, {\it Hyper-K\"ahler metrics of cohomogeneity 
one}, J. Geom. Phys. to appear; E. Calabi, {\it Isometric families of 
K\"ahler structures},
in The Chern Symposium 1979, edited by W.-Y. Hsiang et.al., Springer-Verlag,
New York (1980).

%Two Monopoles TAUB-NUT
\bibitem{Taub} K. Lee, E.J. Weinberg and P. Yi, {\it
Electromagnetic duality and SU(3) monopoles}, Phys. Lett. B376, 97
(1996), hep-th/9601097; J.P. Gauntlett and D.A. Lowe, {\it Dyons and
S-duality in N=4 supersymmetric gauge theory}, Nucl. Phys. B472, 194
(1996), hep-th/9601085; 
S.A. Connell, {\it The dynamics of the SU(3)
charge (1,1) magnetic monopoles,} University of South Australia
preprint.

\bibitem{Rossi}
P. Rossi, {\it Propagation of functions in the field of a monopole}, 
Nucl. Phys. B149, 170 (1979); B.J. Harrington and H.K. Shepard, {\it 
Periodic Euclidean solutions and the finite-temperature Yang-Mills gas},
Phys. Rev. D. 17, 2122 (1978); D. Gross, R. Pisarksi and L.G. Yaffe, 
{\it QCD and instantons in finite temperature}, Rev. Mod. Phys. 53, 
43 (1981).

%Massless monopoles
\bibitem{klee2}
K. Lee, E.J. Weinberg and P. Yi, {\it Massive and massless monopoles with
non-Abelian magnetic charges}, Phys. Rev. D54, 6351 (1996),
hep-th/9605229. 

%Gibbons n-monopole generalization
\bibitem{Gibbons}
G.W. Gibbons, {\it The Sen conjecture for fundamental 
monopoles of distinct types}, Phys. Lett. B382, 53 (1996),
hep-th/9603176. 

\bibitem{Nahm}
W. Nahm, {\it A simple formalism for the BPS monopole}, Phys. Lett. 90B, 413
(1980). 


%Nahm formalism
\bibitem{Nahm5}
W. Nahm, {\it Selfdual monopoles and calorons},  in 
Group Theoretical Methods in Physics, ed. G. Denardo, et.al.,
Springer-Verlag (1984). 




\end{thebibliography}
\end{document}